# Statistical analysis of the limitation of half integer resonances on the available momentum acceptance of a diffraction-limited storage ring

Yi Jiao*, Zhe Duan

Key Laboratory of Particle Acceleration Physics and Technology, Institute of High Energy Physics,
Chinese Academy of Sciences, Beijing 100049, China

*corresponding author. Email address: jiaoyi@ihep.ac.cn

**Abstract**：

In a diffraction-limited storage ring (DLSR), the momentum acceptance (MA) might be limited by the half integer resonances (HIRs) excited by focusing errors, associated with the large detuning terms from the strong focusing and strong sextupoles required for an ultralow emittance. Taking the High Energy Photon Source (HEPS) as an example and through statistical analysis, we found that the horizontal HIRs have stronger impact on dynamics than the vertical ones; and the probability of MA reduction caused by a HIR is closely correlated with the level of the beta beats at the same plane, but independent of the error sources. For the HEPS design, to reach a small MA-reduction probability of about 1%, the rms amplitude of the beta beats at the nominal tunes should be kept below 1.5% horizontally and 2.5% vertically. The presented analysis can provide useful reference for other DLSR designs.

## 1. Introduction

In storage ring light sources, the dynamic aperture (DA) and momentum acceptance (MA) specify the betatron and synchrotron amplitude within which particle motions are safely bounded. They are among the major indicators of the ring performance, as they determine the injection efficiency and beam lifetime, and affect the availability and stability of the photon beams to user experiments.

Nevertheless, to achieve large DA and MA is generally in conflict with the pursuit of low beam emittance and high brightness. To minimize the emittance, strong quadrupoles and sextupoles are needed. The strong fields, together with the machine imperfections, will cause detuning effects (tune shifts with amplitude and momentum deviation) and excite resonances from low to high order, leading to orbit diffusion and even unstable motions. Integer resonances (IRs) and half integer resonances (HIRs) are induced by linear field errors, whose widths depend on the setting of the tunes and the level of the linear field errors but are independent of the betatron amplitudes of particles [1]. While higher order resonances are driven by nonlinear fields (e.g., sextupole fields) or nonlinear field imperfections, whose strengths are generally weak neighboring the ideal particle trajectory, but grow rapidly with increasing amplitude. The major resonances limiting DA and MA may vary among different ring designs. Thus, when designing a storage ring light source, it is essential to identify the resonances that destroy the beam dynamics and to correct them, so as to simultaneously obtain a lowest possible emittance and large enough DA and MA.

In the third generation light sources (TGLSs [2]) widely built around the world, experience [3-4]

indicates that through suitably arranging the sextupoles along the ring and delicately tuning the sextupole strengths and the nominal tunes, the sextupole-induced aberrations can be greatly cancelled or minimized, resulting in small resonance driving terms and detuning terms. The betatron tunes can be kept away from IRs and HIRs even for a large betatron amplitude or momentum deviation, and the beam dynamics is usually dominated by higher order resonances. Nevertheless, it is generally feasible to simultaneously obtain a large DA (typically tens of millimeters) and a large MA (typically larger than 3%) in a TGLS.

However, in a diffraction-limited storage ring (DLSR) with a natural emittance one or two orders of magnitude lower than available in a TGLS, the situation tends to be less optimistic. To reach an ultralow emittance, the double-bend achromat or triple-bend achromat lattices that are commonly used in TGLS designs are no longer desirable; and instead, multi-bend achromat (MBA) lattices are usually used in DLSR designs [5]. Furthermore, novel design philosophies (e.g., the so-called 'hybrid' MBA [6]) and small-aperture magnets [7] are adopted to make the DLSR design more compact and cost effective. On the other hand, since even stronger focusing is used in a DLSR, the natural chromaticities and sextupole strengths are much larger than those in a TGLS. Even with the most advanced analytical (e.g., Lie algebra [8], global or local nonlinearity-cancellation approaches [9, 6]) and numerical optimization techniques (e.g., multi-objective genetic algorithm [10], frequency map analysis [11]), it is extremely difficult to reduce the resonance driving terms and detuning terms to a sufficiently small level. Due to the large detuning terms, the resonances near the nominal tunes are usually reached for small betatron amplitudes or momentum deviations. Therefore, the higher order resonances typically have small strengths and just weakly impact the beam dynamics; while the IRs or HIRs have strong effects on the dynamics, causing small DA and MA and great difficulty in the nonlinear optimization.

In a DLSR, even with only the bare lattice (i.e., without considering any machine imperfection), the DA is typically of the order of 1 mm. It is hard to meet the DA requirement of the off-axis local-bump injection that is mostly adopted in TGLSs. To overcome this problem, scientists proposed methods to inject the beam on-axis into the ring, such as the 'swap-out' [12] and longitudinal injection schemes [13, 14]. It is worthy to mention that it is still possible to attain a DA of up to 10 mm in a DLSR, if using one or more high-beta sections dedicated to off-axis injection [9]. However, the inserted high-beta sections will break the ring periodicity, especially for off-momentum particles, leading to greater difficulty in the MA optimization.

On the other hand, small MA will lead to low Touschek lifetime and poor ring performance, based on the fact that the Touschek lifetime scales as $\delta_m^{3\sim5}$ in a DLSR [9, 15], with $\delta_m$ being the value of the MA. And till now there have been no effective methods to remedy this problem without paying a price in other aspects of ring performance. For example, the Touschek effect can be mitigated to some degree by enlarging the bunch length of the electron beam [16], which, however, will reduce the peak brightness of the emitted photon beam. What is more, a large MA (e.g., $\delta_m >$ 3%) is essentially required if on-axis longitudinal injection is adopted to realize beam accumulation in a DLSR.

As mentioned, the beam dynamics of a DLSR is usually limited by the IRs and HIRs. It is generally thought that the HIRs are less destructive to dynamics than the IRs. Therefore, if the HIRs can be safely crossed, the limitation caused by the HIRs will be removed and the MA optimization will become much easier. Actually, experience [e.g., 17, 18] indicated that the HIRs can be approached or even crossed without beam loss in a TGLS. And a recent simulation study [19] suggested that the HIRs could possibly be crossed without particle loss in a DLSR in the presence of machine imperfections.

Therefore, it is necessary and interesting to find whether and under what conditions the HIRs are not fatal to the dynamics of a DLSR. To this end, taking the High Energy Photon Source (HEPS) design as an example, we statistically analyzed the HIR effects in the presence of different kinds of focusing errors. The probability of MA reduction due to HIRs was found to be positively correlated with the amplitude of the beat beats at the nominal tunes, while not with the source of the beta beats. In addition, to avoid possible MA reduction due to HIRs, the amplitude of the beta beats should be controlled to a sufficiently small level.

The paper is arranged as follows. In Section 2, we will first briefly describe the HEPS design and the MA tracking results with the bare lattice. Then, a model-independent analysis of the probability of MA reduction due to HIRs will be presented in Section 3. And conclusions will be given in Section 4.

**2. HEPS design and tracking results with the bare lattice**

HEPS, a kilometer-scale storage ring light source with a beam energy of 5 to 6 GeV, has been proposed for a few years and is to be built in a near future. The lattice design has been continuously evolved [20-22], along with development of the accelerator hardware systems and progress in the DLSR-related beam dynamics studies.

Recently a hybrid 7BA design with a natural emittance of 60.1 pm·rad at 6 GeV was developed [23] for the HEPS. This design consists of 48 identical 7BAs and has a circumference of 1295.6 m. Each 7BA is about 27 m, with a 6-m straight section for insertion devices. The layout and optical functions of a single 7BA are shown in Fig. 1. In each 7BA, high-gradient (up to 80 T/m) quadrupoles were located neighboring the inner three combined-function dipoles to achieve an ultralow emittance as well as a compact layout. Four outer dipoles with longitudinal gradients were used to create two dispersion bumps with all the sextupoles therein for an efficient chromatic correction. Totally six sextupoles were used in each 7BA and grouped in three families. Between each pair of sextupoles, a –*I* transportation, with phase advance at or close to $(2n+1)\pi$ (where *n* is an integer) in both *x* and *y* planes, was designed to cancel most of the nonlinearities induced by sextupoles. A family of octupoles was also placed in the dispersion bumps and used to reduce the detuning terms. Among these multipoles, since two sextupole families were required for chromaticity correction, only two free knobs were left for nonlinear optimization. This enabled us to perform a grid scan of the nonlinear performance with respect to the multipole strengths in a reasonably short time, based on numerical tracking with the AT program [24] and frequency map analysis. In the tracking, the bare lattice was used, and the radiation effects and synchrotron motion were not considered by assuming a constant momentum deviation (denoted by $\delta$ hereafter).

Unfortunately, it was found difficult to simultaneously obtain a large DA and a large MA (see Ref. [23] for more details).

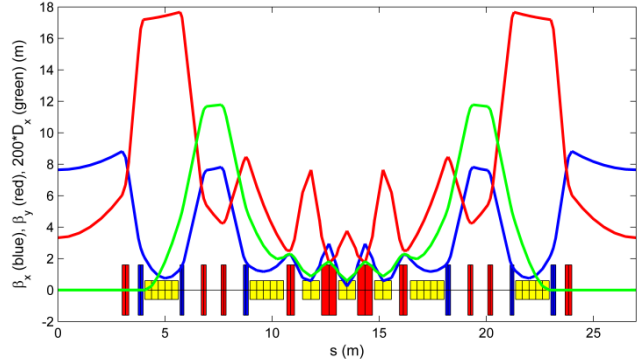

Fig. 1. Layout and optical functions of the hybrid 7BA designed for the HEPS.

The MA and the corresponding frequency map (FM) of the bare lattice of a compromise solution are presented in Fig. 2. And the tune shifts with $\delta$ are shown in Fig. 3. The horizontal HIR $2\upsilon_x = 227$ and the vertical HIR $2\upsilon_y = 83$ are reached for $\delta$ of $-2.95\%$ and $\pm 2.45\%$, respectively. The HIRs cause obvious but moderate distortions in the FM. Particle motions remain stable neighboring the HIRs with confined orbit diffusion. One can see that the coupling resonance $2\upsilon_x - 2\upsilon_y = 48 \times 3$ also has a strong impact on the beam dynamics, but it only affects the horizontal plane.

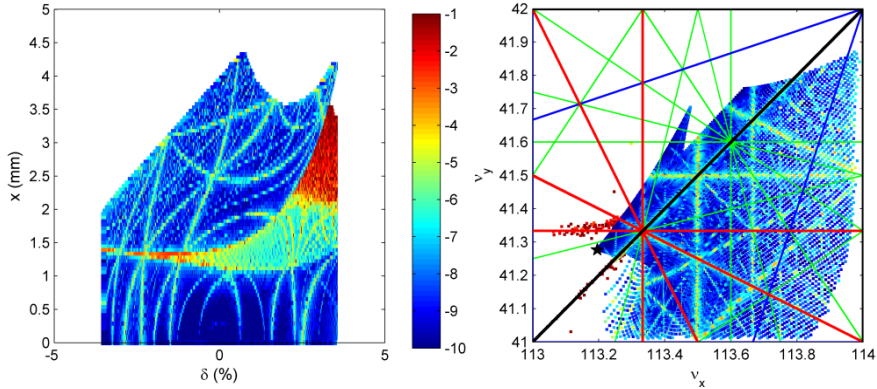

Fig. 2. MA (left) and FM (right) of the HEPS bare lattice. The colors, from blue to red, represent the stability of particle motion, from stable to unstable. In the tracking, $\delta$ was scanned from -3.5% to 3.5%, and the vertical offset was fixed to be 1 $\mu$m.

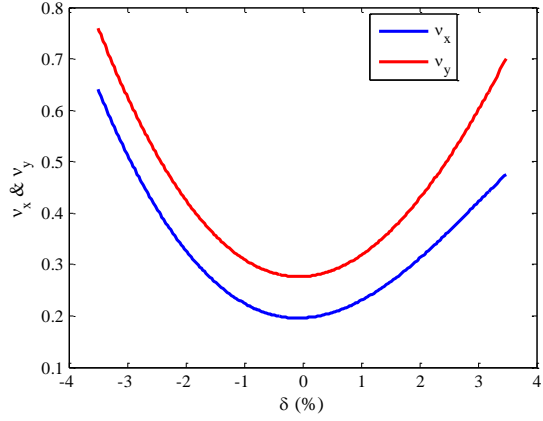

Fig. 3. Tune shifts with $\delta$ of the HEPS bare lattice. In the tune calculations, the transverse offsets were fixed to be 1 $\mu$m.

## 3. Statistical analysis of the HIR-induced MA reduction in the presence of focusing errors

In an electron storage ring, beta beats are caused by focusing errors, mainly from the quadrupole field errors, horizontal displacements at sextupoles and momentum deviations. For the case of particular interest where the nominal tunes are far enough away from the HIRs while the tunes of an off-momentum particle approach the HIRs, the beta beats at a specific $\delta$ can be described by

$$\frac{\Delta\beta(s;\delta)}{\beta(s;\delta)} = \frac{1}{2\sin 2\pi\upsilon(\delta)} \oint \beta(s_1;\delta)\Delta K(s_1;\delta)\cos[2\pi\upsilon(\delta) - 2\Delta\mu(s,s_1;\delta)]ds_1, \qquad (1)$$

where the parameters $\beta$, $\upsilon$, $K$, $\Delta\mu$ are all functions of $\delta$; $\beta(s; \delta)$ and $\beta(s_1; \delta)$ are the beta functions at the longitudinal position $s$ and $s_1$, respectively; $\upsilon(\delta)$ indicates the horizontal or vertical tune of the ring, and gives the nominal tune with $\delta = 0$; $\Delta\mu(s, s_1; \delta)$ is the phase advance between $s$ and $s_1$; and $\Delta K(s_1; \delta)$ is the focusing error, which is of the form

$$\Delta K(s_1;\delta) \approx \Delta K_{qs}(s_1;0)(1-\delta) - K_n(s_1;0)\Delta\delta, \qquad (2)$$

where $K_n(s_1; 0)$ is the nominal focusing strength, and $\Delta K_{qs}(s_1; 0)$ represents the focusing error from a small deviation in the quadrupole field or a nonzero horizontal displacement at sextupole, with $\Delta K_{qs}(s_1; 0) \ll K_n(s_1; 0)$, $\Delta\delta$ is a small deviation on top of $\delta$, with $\Delta\delta \ll \delta \ll 1$, and in Eq. (2) only the first significant terms with respect to $\Delta K_{qs}(s_1; 0)$ and $\Delta\delta$ are kept.

Considering the case with $\upsilon(\delta)$ approaching a HIR, from Eq. (1), the beta beats increase with decreasing $\Delta\upsilon = |\upsilon - N(\upsilon) - 0.5|$, with $N(\upsilon)$ being the integral part of the tune, indicating an increasing resonance width and a greater and greater impact of the HIR on beam dynamics. Nevertheless, as shown in Sec. 2, if considering only the bare lattice [only the second term in Eq. (2) is nonzero], particles will not get lost while crossing the HIRs, due to the counterbalance effects from the large amplitude-dependent detuning terms.

On the other hand, when introducing quadrupole field errors or nonzero displacements in the bare lattice [now the first term in Eq. (2) is also nonzero] to excite larger beta beats, it was observed that the beam dynamics can be destroyed by the HIRs, along with particle loss and MA reduction, even after the tunes were recovered. As a demonstration, the MA and FM of the HEPS lattice in the presence of an ensemble of large quadrupole field errors and horizontal displacements at sextupoles are shown in Fig. 4. One can see that both the horizontal and vertical HIRs have large resonance widths and cause significant distortion in the FM, leading to a reduced MA of about 2.2%.

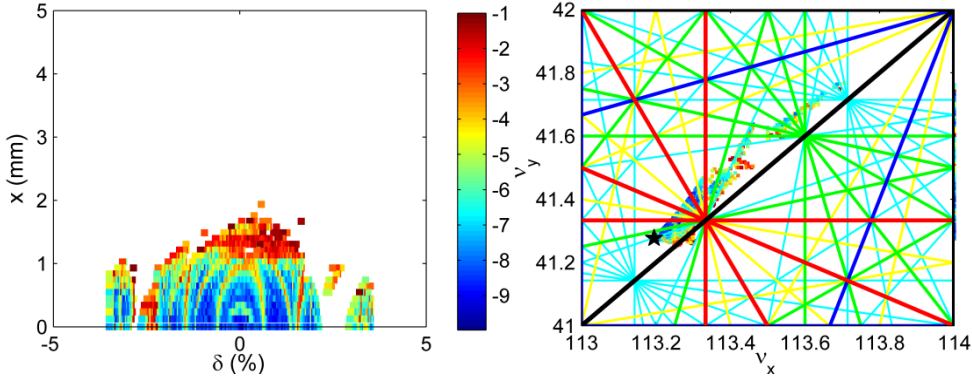

Fig. 4. MA (left) and FM (right) of the HEPS lattice in the presence of an ensemble of large quadrupole field errors and horizontal displacements at sextupoles. The colors, from blue to red, represent the stability of particle motion, from stable to unstable. In the tracking, $\delta$ was scanned from -3.5% to 3.5%, and the vertical offset was fixed to be 1 μm.

To give a quantitative description of the relationship between the possible MA reductions and the focusing errors, we calculated the probability of MA reduction (denoted by $P_{\text{MA, red}}$ hereafter) with respect to the root mean square (rms) amplitude of the beta beats and the rms amplitude of the horizontal displacements at sextupoles with $\delta = 0$, i.e., $(\Delta\beta/\beta)_{\text{rms}}$ and $(x_{\text{DS}})_{\text{rms}}$ at the nominal tunes. The latter parameters were selected because they are, to a large extent, independent of concrete modeling of the focusing errors and can be directly measured during daily operation of a storage ring light source, and therefore the analysis result can be easily verified by beam experiments.

*3.1. Lattice modeling and tracking setup in the presence of focusing errors*

To concisely model the focusing errors, instead of simulating the complicated error generation and lattice calibration process, we generated a large number of ensembles of quadrupole field errors and displacements at sextupoles, added them into the bare lattice, simply restored the tunes with two families of quadrupoles, and then calculated the $(\Delta\beta/\beta)_{\text{rms}}$ and $(x_{\text{DS}})_{\text{rms}}$. The values of $(\Delta\beta/\beta)_{\text{rms}}$ and $(x_{\text{DS}})_{\text{rms}}$ were used to imitate the residual deviation of the optical parameters after a more thorough correction. Based on the fact that the phase advance between the sextupoles and the 7BA center is about odd integer times of π/2, random horizontal dipole kicks were located at the centers of 7BAs, which transfer to nonzero displacements at the sextupoles.

And then, particles with $\delta$ ranging from -3.5% to 3.5% were launched for tracking over 1000 turns, to obtain the corresponding $\delta_m$. If particles get lost at several $\delta$ values, the $\delta_m$ will be assigned the smallest $\delta$ absolute value; otherwise, the MA will be set to 3.5%.

Instead of assuming a constant $\delta$, the synchrotron motion and radiation effects were included in the tracking. A 100 MHz RF system with a voltage of 2.5 MV was added in the lattice to provide a bucket height of 3.8%. As shown in Figs. 5 and 6 where the same focusing errors as set in Fig. 4 were used and the RF and synchrotron radiation were turned on, in the case with $\upsilon(\delta)$ above a vertical HIR, the particle was trapped by the HIR and eventually got lost within 400 turns. This process lasts a much shorter time compared to the damping time, about 4500 turns. Thus, including the radiation effects in the tracking does not affect the calculated $\delta_m$. Nevertheless, in the simulation results shown in the following subsection, the radiation effects were always considered.

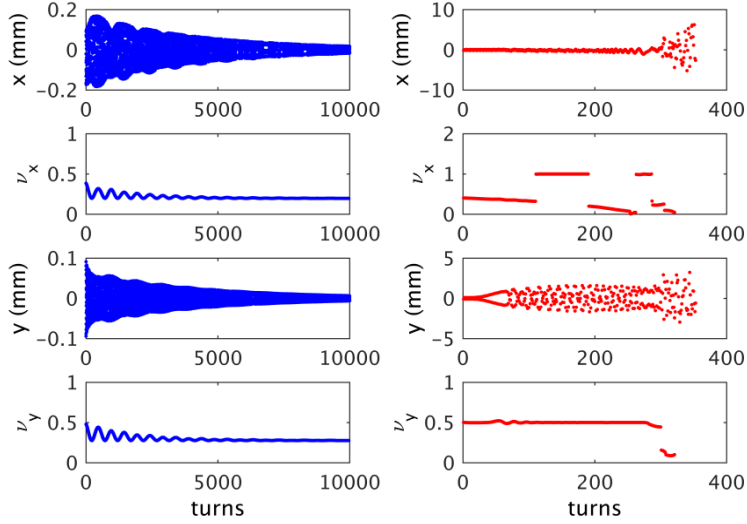

Fig. 5. Evolution of the transverse coordinates and the fractional tunes with initial $\delta$ of -2.35% (blue) and -2.45% (red), which correspond to $\upsilon(\delta)$ below and above a vertical HIR, respectively. The same focusing errors as set in in Fig. 4 were used and the RF and synchrotron radiation were turned on. The initial transverse offsets were set to 0.1 mm. The tunes were calculated with the NAFF analysis [25] of consecutive 32 turns of transverse coordinates.

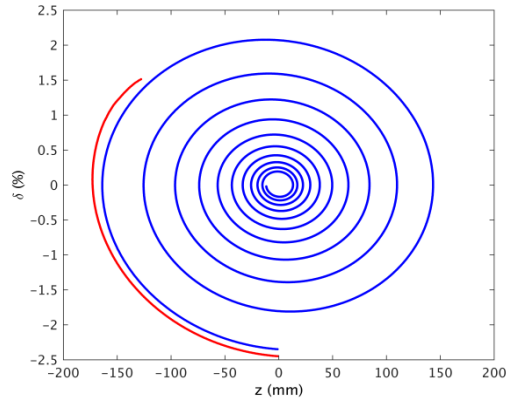

Fig. 6. Synchrotron motions with initial $\delta$ of -2.35% (blue) and -2.45% (red).

### 3.2. Statistical analysis of the MA reduction due to HIRs

First, we considered the simplest case with only the quadrupole field errors in the lattice, with $(x_{DS})_{rms} \equiv 0$. Totally $5 \times 10^4$ ensembles of random errors were generated and evaluated, and the error amplitude was set so that the resulting $(\Delta\beta_x/\beta_x)_{rms}$ and $(\Delta\beta_y/\beta_y)_{rms}$ are less than 10% for most of the error ensembles. The results are presented in Fig. 7. It appears that the MA reduction is more likely to occur for larger beta beats. In addition, for the cases with reduced MA, most of the $\delta_m$ values are around 2.4% or 2.9%, corresponding to particle loss due to the vertical or horizontal HIRs, respectively.

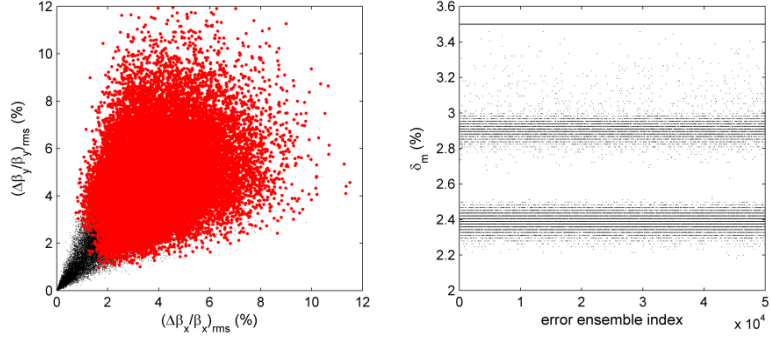

Fig. 7. Cases of MA reduction (red points in the left plot and the points with $\delta_m < 3.5\%$ in the right plot) in the presence of $5 \times 10^4$ ensembles of the quadrupole field errors (all the points).

The HIR effects were further analyzed by measuring the probability of MA reduction $P_{\text{MA, red}}$ with respect to different beta beat amplitudes. For a specific set of $(\Delta\beta_x/\beta_x)_{\text{rms}}$ and $(\Delta\beta_y/\beta_y)_{\text{rms}}$, if there were more than 100 error ensembles in the vicinity, the $P_{\text{MA, red}}$ will be calculated by counting the proportion of the cases with MA reduction. It was found that the $P_{\text{MA, red}}$ caused by a specific HIR relies mainly on the $(\Delta\beta/\beta)_{\text{rms}}$ at the same plane, as shown in Fig. 8. This is particularly pronounced in the cases with $(\Delta\beta/\beta)_{\text{rms}}$ below 3%, which is close to the achievable level of the residual beta beats in TGLSs with the state-of-art optics correction techniques. One can also see from Fig. 8 that to reduce $P_{\text{MA, red}}$ to about 1%, $(\Delta\beta/\beta)_{\text{rms}}$ needs to be kept below 1.5% in the $x$ plane and below 2.5% in the $y$ plane, respectively. This indicates that with the same level of beta beats in the $x$ and $y$ planes, the horizontal HIRs are generally more dangerous than the vertical ones, which accords with the experimental observations in TGLSs [e.g., 17]. This can be understood by recalling that in a storage ring light source strong horizontal focusing is required to attain a low emittance, while vertical focusing can be relatively weaker.

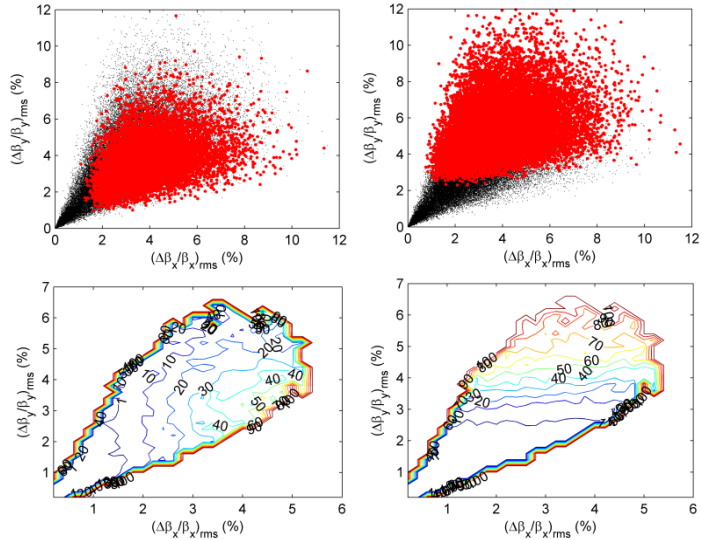

Fig. 8. Cases of MA reduction and contours of the MA-reduction probability (in unit of %) caused solely by the horizontal (left) or the vertical HIR (right).

Secondly, we also considered the random dipole kicks that induce nonzero horizontal displacements at sextupoles. In this case, more error ensembles ($1 \times 10^5$) were generated, and the dipole kicks were set so that the maximum $(x_{DS})_{rms}$ is about 30 μm and the induced $(\Delta\beta/\beta)_{rms}$ is about 10%. Therefore, some error ensembles resulted in $(\Delta\beta/\beta)_{rms}$ larger than those with only the quadrupole field errors. Nevertheless, it was found that including nonzero horizontal displacements at sextupoles does not make any difference in the contours of the $P_{MA, red}$ with $(\Delta\beta/\beta)_{rms}$ below 3%. In addition, as shown in Fig. 9, there is no obvious correlation between the $P_{MA, red}$ and $(x_{DS})_{rms}$, indicating that the HIR effects are not related to the concrete source of the beat beats.

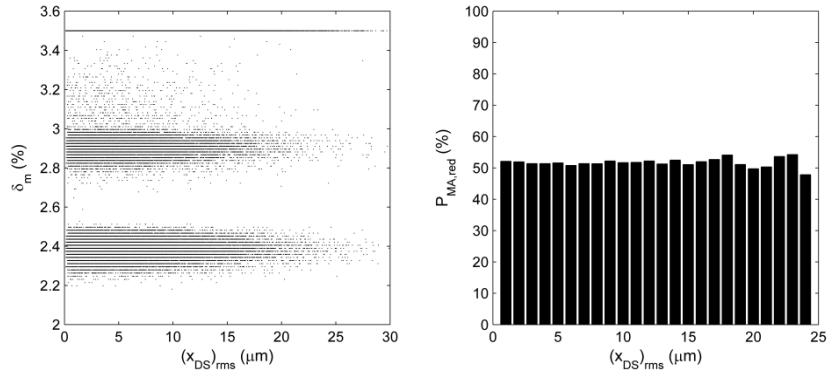

Fig. 9. MA values (left) and the MA-reduction probabilities (right) with respect to the rms amplitudes of the horizontal displacements at sextupoles, for the HEPS lattice in the presence of $1 \times 10^5$ ensembles of the quadrupole field errors and dipole kick errors.

4. Conclusion

In this paper, taking the HEPS design as an example, we statistically analyzed the limitation effects of HIRs on the available MA in a DLSR. Three kinds of focusing error sources were considered, i.e., the quadrupole field errors, horizontal displacements at sextupoles and the momentum deviations. It was found that the HIRs will not lead to MA reduction in the absence of the former two sources.

The contribution of the focusing errors to the HIR effects were represented with the level of the beta beats at the nominal tunes. For the presented HEPS design, it was found that to reach a MA-reduction probability of smaller than 1%, the rms amplitude of the beta beats should be reduced to a sufficiently small level, i.e., less than 1.5% in the $x$ plane and 2.5% in the $y$ plane, respectively. Although the limitation effects of the HIRs on only the MA was investigated in this study, the presented analysis and the obtained criterion are applicable for the limitation effects of the HIRs on the available DA as well.

Apart from the presented HEPS design, other HEPS designs [26] with similar layout but with different emittance and nominal tunes were also tested. Basically the same criterion was obtained for safely crossing the HIRs. This is because although these designs have different parameters from the presented design, they use similar quadrupole and sextupole strengths as required by the emittance minimization, and thus have almost the same response and tolerance to the deviation in optical parameters. Based on the same consideration, the obtained criterion can provide a useful reference to other DLSR designs.

For a more rigorous analysis of the limitation effects of the HIRs on the available MA, one needs to look at the relationship between the MA-reduction probability and the beta beats at a specific $\delta$ with $\upsilon(\delta)$ close to a HIR, rather than those at the nominal tunes. Simulation studies revealed that the amplification factor of the beta beat amplitude in these two cases depends on concrete setting of the focusing errors, but always fluctuates around the value of $\sin[2\pi\upsilon(0)]/\sin[2\pi\upsilon(\delta)]$. Besides, as mentioned above, using the beta beats at the nominal tunes as observation variables, the obtained criterion can be easily followed and verified during both the design and operation of a DLSR.

At last, it is worthy to remark that in the preliminary design stage of a DLSR, a bare lattice is handy before comprehensive error generation and correction could be implemented. Rather than the conventionally defined DA and MA, the authors recommend using the 'effective' DA and MA as indicators of the nonlinear performance of a ring, within which, not only the motions remain stable after tracking over a few thousand turns, but also the betatron tunes are bounded by the IRs and HIRs nearest to the nominal tunes. The effective DA and MA calculated with the bare lattice can provide a quick and reasonable measure of the ring acceptance of a realistic machine, and will facilitate the optimization of a DLSR design.